\documentclass[sigconf,authorversion,nonacm]{acmart}

\AtBeginDocument{%
  }

\usepackage{graphicx} % Required for inserting images
\usepackage{xcolor}
\usepackage{hyperref}
\usepackage{natbib}% Citation support using natbib.sty
\usepackage{amsfonts}
\bibpunct[, ]{(}{)}{;}{a}{}{,}% Citation support using natbib.sty
\usepackage{amsmath}
\graphicspath{{./figures/}}

\def\betab{\mbox{\boldmath $\beta$}}

\def\xb{\mathbf{x}}             % Bold x
\newcommand{\myT}{^\mathsf{T}}  
\newcommand{\Prob}{\textsf{P}}
\newcommand{\E}{\mathbb{E}}
\newcommand{\V}{\mathbb{V}}

\usepackage{soul} % for strikethrough
\newcommand{\stkout}[1]{\ifmmode\text{\sout{\ensuremath{#1}}}\else\sout{#1}\fi} % make the strikeout work for equations

\newcommand{\hypoth}[2]{\par\textbf{#1}: #2\par}

\date{October 27, 2023}

\begin{document} %{

\title{Forecasting Future News Deserts}


\author{Edward Malthouse}
\orcid{0000-0001-7077-0172}
\email{ecm@northwestern.edu}
\author{Jaewon Choi}
\author{Zach Metzger}
\author{Larry DeGaris }
\affiliation{%
  \institution{Medill School, Northwestern University}
  \streetaddress{1845 Sheridan Road}
  \city{Evanston}
  \state{Il}
  \country{USA}
}

\renewcommand{\shortauthors}{Malthouse et al.}

\begin{abstract}
This article builds a model to forecast the number of newspapers that will exist in each US county in 2028, based on
what is known about each county in 2023. The methodology is to use information known in 2018 to predict the number of
newspapers in 2023. Having estimated the model parameters, we apply it to 2023 data. The model is based on market
demographic characteristics and allows for different effects (slopes) for large, medium and small markets (population
segments). While the main contribution is forecasting, we interpret the parameter estimates for validation. We find that the
best predictor of the number of newspapers in five years is the current number of newspapers. Population size
also has a positive association with newspapers. Average age and median income have positive slopes, but not in all
population segments. The proportions of Blacks, and separately Hispanics, in a county have negative associations with the
number of newspapers, but not in all population segments. The report provides maps showing which counties that are currently news deserts could be revived, which counties that currently have one newspaper are more at risk of losing it, and which counties with two or more newspapers are at risk. We also study the model residuals showing which counties are under- or over-performing relative to the market conditions. 
\end{abstract}

\maketitle

%===================================================================
\section{Introduction}

Local journalism fulfills the function of providing ``critical information needs'' of the citizens
\citep{waldmanInformationNeedsCommunities2011} and constructing an informed public, a foundation of American democracy
\citep{gao2020financing,mondak1995newspapers}. The continued decline of local newspapers and the risk of waning local
journalism have been well-documented by a series of studies by Penny Abernathy and her team, first at the University of North
Carolina (UNC) and now at Northwestern University \citep{abernathyNewsDesertsGhost2020a, abernathyStateLocalNews2022a}. They
undertook the laborious task of cataloging existing newspapers, tallying the number of newspapers in
each county of the United States, and creating maps that show which geographic regions have either completely lost coverage
(no newspapers) or have only one newspaper. Their work has documented the current state of the local news ecosystem
\citep{abernathyStateLocalNews2022a} and sounded an alarm having widespread impact, prompting discussion in journalism, policy, philanthropic and academic circles over what to do about the news crisis that is unfolding.
%A news desert is ``a community, either rural or urban, with limited access to the sort of credible and comprehensive news and
%information that feeds democracy at the grassroots level'' (\ecmadd{cite
%\href{https://www.cislm.org/what-exactly-is-a-news-desert/}{UNC} or Medill???}). The definition does not specify a specific
%number of newspapers and when we need this specificity we will give specific count values, e.g., counties with 0 or 1
%newspapers.

This research attempts to extend Abernathy's news desert research by building a forecasting model. A forecast is a prediction
of what will happen in the future, as opposed to a description of the current situation. Forecasts require a statistical or
machine-learning model that takes as inputs information about the past and present, and outputs an estimate of the number of
newspapers in a county in the future. The goals are to develop such a model and communicate the forecasts that it generates.
In particular, we forecast the number of newspapers in each county five years from now (2028) and create visual maps showing the results. We envision this work being of use to all actors concerned about the future of local news, including policy makers, philanthropists, investors, journalists, etc.

Our approach is to predict the current number of newspapers in each county based on the \emph{market conditions} as of five
years ago (2018). The specific market conditions will be discussed below. Having estimated and validated the model, we apply
it to the market conditions as of now (2023) to produce forecasts of the number of newspapers in five years (2028).
%This is a forecasting model rather than a description of what happened. Need to capture timeless causal factors, not specific causes that happened once in the past. 

%The present work builds multivariate models to predict news deserts from \emph{market conditions}, described in detail in section~\ref{ss:mktcond}. In doing so, this work extends the previous work with three goals.
%\begin{enumerate}
%\item
%\label{rq:interpret}
%Explain why counties have become news deserts or ``at risk'' based on their market conditions. Likewise, which market conditions are associated with a ``safe'' news ecosystem?
%\item \label{rq:predict}
%How accurately can we predict the future news landscape in a county? Our models will provide predictions of where news deserts are likely to arise in the future, which will be called \emph{at risk} counties. Conversely, the model may identify areas that are currently news deserts that should not be given the market conditions. Such news-desert counties would seem to be prime targets that could most easily be revived. The models will also identify \emph{safe} counties, which are not news deserts and unlikely to become ones in the next few years.
% \item \label{rq:map}
%We visualized the results of the model with a map showing which counties are likely to become news deserts, along with those that are currently news deserts or safe.
%\end{enumerate}

%===================================================================
\section{Literature and Conceptual Framework}

Research on news deserts is in a burgeoning stage and current studies tend to
be simple descriptions at the macro-level
\citep{claussenDigestingReportNews2020}, or thick descriptions at the
micro-level \citep{mathewsLifeNewsDesert2022,mathewsDesertWorkLife2022}. This
study advances the discussion by providing explanations and predictions at
the macro-level. A few empirical studies include demographic community
characteristics in their models. For instance,
\citet{choInternetAdoptionSurvival2016} investigated the impact of broadband
penetration rate on the number of newspaper titles at the national level.
\citet{sarah23} documented news coverage in New Jersey and correlated it with demographics. 

The work most closely related to ours is \citet{napoliAssessingLocalJournalism2018}, which studied the relationship between
demographic and geographic characteristics and ``local journalism robustness.'' Although our research goals, research design
and measures are different, we use their work as a foundation, and build from it to create our contribution. Sticking close to their design and corroborating their findings bolsters the credibility of our model. We first summarize their research design and findings. Next, we extend their model with additional covariates and population segments, and justify our modifications with relevant literature.
 
\citet{napoliAssessingLocalJournalism2018} studied ``communities'' (their sampling unit) with populations between 20,000 and 300,000. They identified 493 such communities in the US and drew a random sample of 100 for their study. They audited local newspapers, local radio stations, local TV stations and local online-only news for stories about the communities. Their audit measured four dependent variables (DVs): counts of news stories produced for the community, original stories, local news stories, and critical-information-need stories \citep{waldmanInformationNeedsCommunities2011}. Their DVs thus quantified the \emph{amount} of different types of news coverage, which they term \emph{journalism robustness}. They profiled news coverage across communities and fitted a regression model predicting the four story counts. Their regression models, in particular, motivate our approach.

Their predictor variables and effects (estimated slopes) on the amount of news coverage were as follows. \textbf{Population} was the only variable that had highly significant effects for all four DVs.
They noted that universities often create local news stories and found the \textbf{number of universities} to have a significant positive slope with three of the counts. 
The \textbf{distance} from a top-50 metro area had a significant positive effect on two DVs, and positive but non-significant effect on the other two. This suggests that the further a community is from a large metro, the more local news coverage, perhaps indicating gaps in coverage of suburbs and exurbs.
The \textbf{percentage of Hispanics/Latinos} had significant negative associations with two of the counts and negative but non-significant slopes for the others. 
None of the other predictor variables were significant, including median household \textbf{income}, 
the \textbf{percentage of African Americans}, whether the community is a county seat, and \textbf{population density}. 
In contrast, \citet{sarah23} found positive population density\footnote{To explain the difference in population effects, Napoli uses a multivariate model, which likely has multicollinearity between population and population density, while Stonbely uses a crosstab with density, but not population size.}, income and Hispanic effects, and a ``suggestive but not statistically significant'' negative effect for African-Americans. These conflicting findings highlight the need for additional research and triangulation with different research designs, which we will provide.
%They did not perform any post-hoc variable selection, and their primary goal was explanatory as opposed to forecasting.

\begin{figure}[htb]
    \centering
    \includegraphics[width=.45\textwidth]{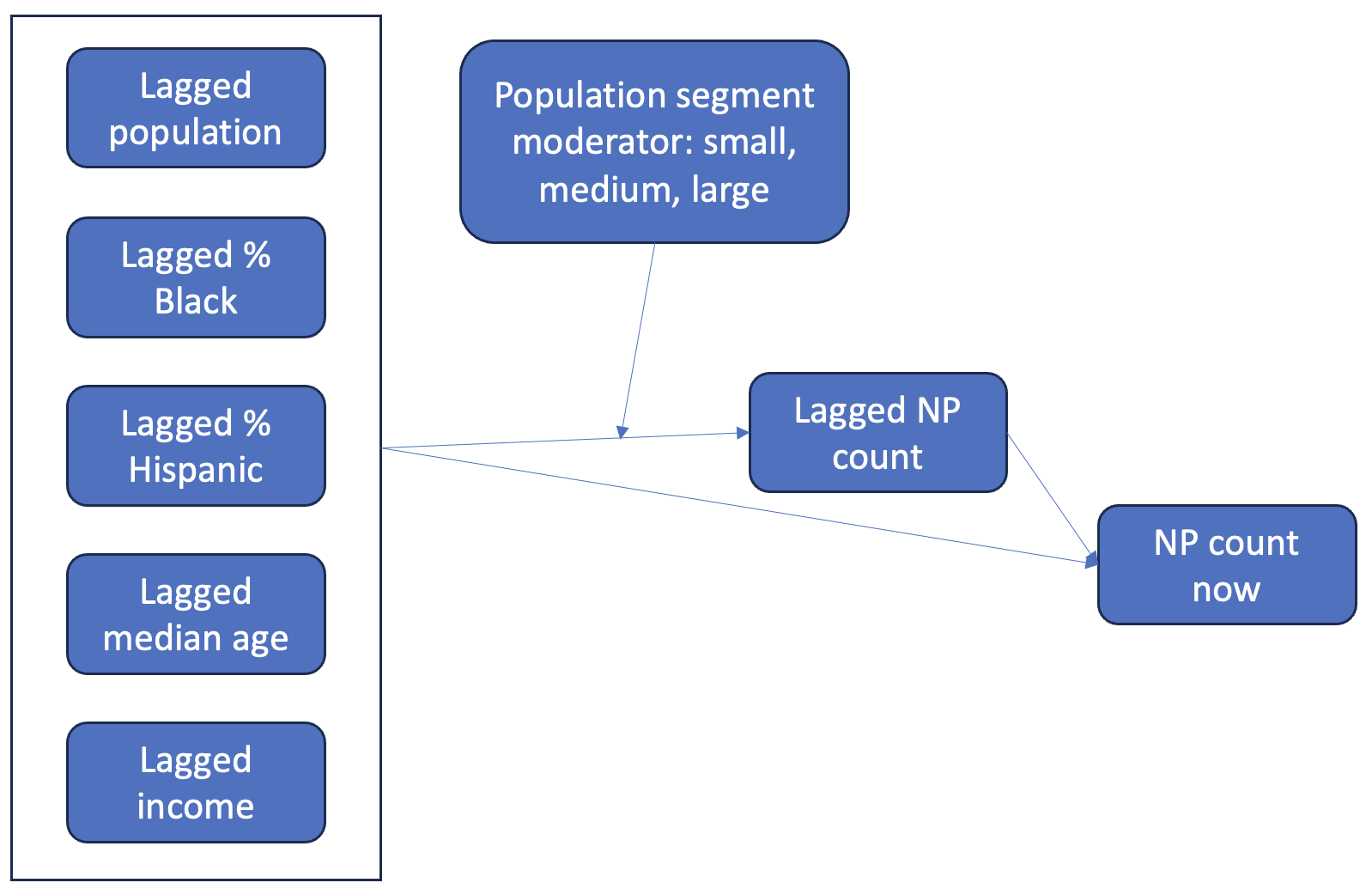}
    \caption{Conceptual framework for our model}
    \label{fig:framework}
\end{figure}

Figure~\ref{fig:framework} shows our conceptual framework, which builds upon
\citet{napoliAssessingLocalJournalism2018} by adding two variables and
dropping others due to other differences in our research design. We begin by
discussing variables in Napoli's model that we dropped. The first difference
is that our unit of analysis is the county rather than community. Since each
county has a county seat we exclude that dummy. A second difference is that
we study the number of local newspapers in the county rather
than story counts. While universities often create local stories, they do not
create newspapers; thus we drop the number of universities as a predictor. A third difference is that, in addition to the population segment of 20K--300K, we will also study counties with population greater than than 300K and those with population less than 20K. We call these \textbf{population segments}. We allow for market conditions to have different slopes in different population segments, making it a moderator. Aligning our population segments with Napoli's allows for a more direct comparison of the findings. Population and population density are highly correlated, and both are correlated with urbanicity and distance from large metros. We decided that only one of the three could be included in the model to avoid multicollinearity, and selected population, since it was the dominant predictor in \citet{napoliAssessingLocalJournalism2018}. Multicollinearity is not a major concern when prediction, as opposed to interpretation, is the model goal \citep{liu2017primer}, but as mentioned in the introduction, we are willing to sacrifice incremental improvements to accuracy for parsimony and transparency. As robustness checks, appendix~\ref{ss:rucc} reports models replacing population with the US Department of Agriculture (USDA) RUCA (rural-urban commuting area) codes.

We develop formal hypotheses to explain the existence of newspapers. Note that newspapers can generate the
revenues required to sustain operations from different sources including advertising; reader revenue from subscription fees,
memberships or donors; and philanthropy. The availability of these potential revenue streams should have a positive effect on
the number of newspapers. Because counties with larger populations have a larger potential market for subscribers/members and for advertisers, we expect
\hypoth{H1}{Counties with larger populations have more newspapers.}
\noindent Likewise, more affluent\footnote{More broadly, socioeconomic status
(SES), which goes beyond income and includes other indicators such as
education levels. We focus on household income to be consistent with
\cite{napoliAssessingLocalJournalism2018}, and report a model in
appendix~\ref{ss:ses} using a composite SES variable instead of income as a
robustness check. Appendix~\ref{ss:poverty} reports another model using poverty instead of income.} counties will have have more money to support businesses (advertising), more willingness to pay for news, and more philanthropy. \citet{hamilton2018poor} discusses the implications of poor local news information in low-income areas, bolstering the importance of this hypothesis.
\hypoth{H2}{Counties with higher median household income will have more newspapers.}

There are several reasons why Black and Hispanic-majority communities may have fewer newspapers. First, different ethnic groups have distinct communication cultures, indicating different patterns of news consumption. For instance, Black communities tend to prefer TV as opposed to whites who prefer newspapers \citep{beaudoinExploringAssociationNews2009,naseerTrendsFactsHispanic2023}. Such historical and cultural trends imply that counties with majority Black populations might not have been seen as desirable markets for newspapers in general, despite their higher attentiveness to local news stories \citep{mitchellLocalNewsDigital2015}. Furthermore, it has been suggested that patterns in the local journalism availability reflect the patterns of digital divide \citep{napoliLocalJournalismInformation2016,napoliLocalJournalismAtrisk2020}, implying that Black and Hispanic communities may have less access to robust local journalism. For these reasons we hypothesize
\hypoth{H3}{There will be a negative relationship between the percentage of Black residents and the number of newspapers.}
\hypoth{H4}{There will be a negative relationship between the percentage of Hispanic/Latino residents and the number of newspapers.}

Furthermore, we consider the median age of the county. Studies have found a positive association between age and reading
newspapers \citep{loges1993dependency, burgoon1980predictors,malthouse2006demographics}. More recent investigations
corroborate that older populations are more likely to read and prefer print newspapers than younger generations \citep{forman-katzNewsPlatformFact2022,wadbringPrintCrisisLocal2017,zerbaYoungAdultsReasons2011}. In addition to such a disposition, older population tends to have a lower level of digital proficiency despite their increased adoption and use of digital/online technologies \citep{hargittaiOldDogsNew2017, hunsakerReviewInternetUse2018,quan-haaseDividingGreyDivide2018}. Given their preference and relative lack of digital capabilities, older counties are more likely to remain desirable markets for local newspapers, who still rely heavily on print circulation. This leads us to hypothesize that
\hypoth{H5}{Age has a positive effect on the number of newspapers.}

Finally, we also account for the past health of the news ecosystem. A county that had many newspapers in the past is less
likely to become a news desert than a county that only had one newspaper in the past because of the brand equity. The
existence of newspaper brands create momentum that will help sustain the enterprise. We refer to this as the \textbf{lagged
number of newspapers}. This variable is pivotal for prediction because it carries more complicated nuance. While market
demographics are fairly stable over short time periods (e.g., 5--10 years), the lagged number of newspapers may reflect a
consequence of the market conditions that continued for longer periods of time. In the language of casual models
\citep{pearl2018book}, the lagged number of newspapers forms a ``pipe'' between market conditions and the number of newspapers in 2023. Note that when interpreting the effects of market conditions on the current number newspapers, the lagged count should \emph{not} be in the model because it ``blocks the pipe'' and leads to incorrect inferences. For prediction models, the lagged count should be in the model.

Including lagged newspapers as a predictor accounts for unobserved variables. For example, suppose that there is an unobserved
factor such as culture, where some counties are more supportive of their local newspapers than others. This unobserved variable will have caused the lagged number of newspapers, and a model that includes lagged newspapers will indirectly account for the unobserved variable. 

One difficulty with lagged number of newspapers is in selecting how far back to lag. We can consider some extreme cases to elaborate. For instance, a one-day lag (number of newspapers in the county yesterday) would be too short because it would be equal to the dependent variable for all counties except those that had a newspaper closing (or startup) in the past day. Conversely, lagging by 20 years is too long because the technology, revenue and ownership models have changed substantially over this period, e.g., shifts to digital technologies, shifts from advertising to reader and philanthropic supported business models, and the emergence of private equity and regional baron ownership models. We choose five years lag because it was past 2008 global financial crisis, had similar technological environment, and yet offers considerable time period for some major changes in the industry. 
%Do we need to mention how far back we decided to go and justification for it here? I added some since it seemed natural to provide such details.

%Penny said that there are four main indicators in the literature:
%\begin{itemize}
%    \item Population density
%    \item Affluence (SES)
%    \item Ownership
%    \item Owner and community investment
%\end{itemize}

%===================================================================
\section{Research Design and Methods}

The overall design is to predict the number of newspapers as of 2023 using data on the market conditions as of 2018.
To predict which counties will become news deserts in the future,
we apply the model to the market conditions as of 2023.

%------------------------------------------------------------
%\subsection{Data collection}

%---------------------------------
\subsection{Modeling Approach}

Many modeling decisions will have to be made and we now discuss some overall guiding principles. One decision concerns how interpretable the model should be. There are many machine-learning approaches that treat the model relating the inputs to the output as a black box \citep{breiman2001statistical}: the goal of such models is to forecast the future as accurately as possible; the way that inputs affect the outputs is unimportant (black box). As a concrete example, suppose that the median household income in a county affects the number of newspapers in the future. A black-box, machine-learning approach (e.g., neural network or random forest) would allow for income to affect the forecast, but it would be complicated to understand exactly how income affects the outcome, such as whether the association is positive or negative, linear or nonlinear, or whether the effect of income on newspapers changes depending on the values of other inputs in the model (interaction terms), e.g., income might have a stronger effect in small communities than large ones. On the other end of the continuum are linear models, which have the virtue of being transparent, parsimonious and easy to explain. With a linear model, income would appear in a linear equation and we can transparently say that every increase in income of \$1000 is associated with some fixed increase (the slope) in the number of newspapers, on the average. A possible limitation of these models is that their forecasts may not be as accurate as those from the machine-learning methods. We anticipate that our model will be closely scrutinized and therefore we are willing to sacrifice incremental amounts of prediction accuracy for parsimony and transparency. Furthermore, we will be able to compare our estimated slopes with other published research studies as an additional way of validating our model. We will compare our linear model to machine-learning approaches to quantify the reduction in accuracy.

Another modeling decision concerns the inclusion of
covariates. Many forecasting models do not include covariates and instead
rely only on previous values of the criterion, in our case the number of newspapers in a county. The simplest of such methods is called the na\"ive
(or stationary) forecasting model \citep[\S~3.1]{hyndman2018forecasting}, which forecasts the number of newspapers in the next period simply as the number that exist today. We argue that covariates are essential because the market is changing rapidly. For example, 20 years ago newspaper derived a large share of revenue from advertising and the ownership models were mostly privately and publicly held companies. Since then other companies have claimed a large share of the advertising spend (e.g., Craigslist and others for classifieds, Google and Facebook, etc.), and many other ownership models have been introduced, e.g., private equity firms buying large numbers of local newspapers. The factors that made a newspaper viable in the past may not be the same as those today, and we expect non-stationarity. The recent past has been further disrupted by a global pandemic and other world events, moving us far from any ``steady state.'' Forecasts should therefore be based on a broader set of factors beyond the current number of newspapers and these factors (covariates) should have a strong theoretical rationale for why they should affect the viability of newspapers in the future. 

%---------------------------------
\subsection{Dependent Variable and Model}

A simple and natural measure of the health of a news ecosystem is the number of daily and weekly newspapers in the county,
hereafter referred to as the \textbf{count}. The count variable distinguishes between counties with no newspapers, one
newspaper, two newspapers, versus counties with healthy ecosystems consisting of dozens of newspapers. It thus describes the entire range, from being a total news desert through very robust. %With such a large amount of variation, we expect the count to have the most statistical ``power'' in detecting significant predictor characteristics. We are consistent with \citet{napoliAssessingLocalJournalism2018} in modeling a quantity variable. 
Our measure is constructed using various sources. The underlying dataset of newspapers has been in development since 2005 and is drawn from the membership directories of each state’s press association. Additional newspapers have been incorporated on a case-by-case basis after manual review. Deliberately excluded from this dataset are papers that do not provide substantial local news coverage, including shoppers, specialty publications such as business journals and lifestyle magazines, monthly and bimonthly publications, and advertising inserts.
%The extent to which a county is a news desert can be measured in different ways. This study presents three approaches, which are defined and discussed in this subsection. It is necessary to examine the three approaches because we will argue that, while they should largely tell a consistent story, they have different strengths and weaknesses, lending themselves to answering different questions.

%\subsubsection{Counts}
\newcommand{\countmodel}{Poisson}

We use a \countmodel\ model with a log link function, and so the model predicts the log mean newspaper count in each county. For county $i=1, \ldots, n$, let $y_i$ be a realization of random variable $Y_i$ giving the observed number of newspapers, and $\xb_i$ be a $p$-dimensional vector of predictor variables, i.e., the market conditions discussed in the next subsection. The Poisson model assumes that $Y_i$ has a Poisson distribution with mean $\E(Y_i)=\mu_i$ that is related to its market conditions as follows:
\[ \log(\mu_i) = \betab\myT \xb_i, \]
where $\betab$ is a $p$-vector of slope coefficients.

%------------------------------------------------------------
\subsection{Market Condition (Independent) Variables}
\label{ss:mktcond}

%................................................
%\subsubsection{Census data and demographics}
We use the American Community Survey (ACS) five-year estimate data provided by the Census Bureau through its API. The demographic and socioeconomic status variables relevant to our hypotheses are retrieved. These include total population, median age, median household income, and racial/ethnic characteristics of the county. 
%Population density is calculated as population per square kilometer. 
Following \citet{napoliAssessingLocalJournalism2018}, we use the \emph{proportions} of Black and Hispanic populations, respectively, as indicators of racial/ethnic characteristics of the county.
%Additionally, we incorporate socioeconomic status (SES) variable in our analysis. The level of affluence and education is expected to be closely related to local newspapers' survival. Median household income, the percentage of adult population with educational attainment of Bachelor's degree or higher, and the percentage of adult population with educational attainment of less than high school are retrieved from the ACS database. As educational attainment and household income levels are highly correlated, we constructed a composite variable using principal component analysis. Population density, percentage of Hispanic, and percentage of African American variables are log-transformed due to skewness.
%................................................

%\subsubsection{Ownership}

%\ecmadd{(Zach tells how he collected the ND data)}

%---------------------------%------------------------------------------------------------
%\subsubsection{Lagged newspapers}

For the variable reflecting the past health of the news ecosystem, we use the number of newspapers in each county as of 2018,
which we will call \emph{lagged newspapers}.
%This should be an important predictor variable because a county that has many
%newspapers has more chances of having newspapers several years from now. In contrast, a county that only has two newspapers will have fewer ways of remaining safe. 
This variable should be included when assessing how accurately the model can forecast the future number of newspapers. Including lagged newspapers alone is an autoregressive model of order 1, or ``AR(1),'' \cite[\S8.3]{hyndman2018forecasting}, and is a widely-used forecasting model that we will use as a benchmark.

\subsection{Model Validation}
We validate the model in two ways. First, to show consistency with \citet{napoliAssessingLocalJournalism2018}, we re-estimate
the model without lagged newspapers as a predictor. Second, we hold out a test set and compare the predictions of our model
with other benchmarks, showing that our simple, parsimonious model performs roughly as well as more complex models and that
the market condition variables are necessary. The unit of analysis will be the county because that is the lowest level of
geography that is currently available in the UNC/Northwestern news desert data. 
%The limitations of this unit of analysis are acknowledged and discussed in future research section~\ref{ss:futureResearch}.
%Based on the predicted probabilities from our model, we create maps to visualize the expected state of local news ecosystem.

%===================================================================
\section{Results}

We make our data and predicted values available. Our code and output are also available as an R Markdown file.

%------------------------------------------------------------
\subsection{Descriptive Statistics}
\label{ss:desstat}

We report descriptive statistics to provide a better understanding of the data set. Table~\ref{tab:desstat} reports univariate descriptive statistics of the variables included in the model. All variables except for age and income are highly right skewed and we apply log transforms\footnote{The percentage of Black and Hispanic population both take the value 0 and an offset of 1 was added before taking the log to avoid logging 0, which is undefined.} to symmetrize their distributions, increase the density of points in the right tail, and reduce the influence of extreme values. The last three rows show descriptive statistics for the transformed variables, and skewness is greatly improved. The dependent variable is the number of newspapers in 2023 and there are 204 counties with zero newspapers, 1,628 with one, 629 with two, 293 with three, \ldots, one (Cook County, Illinois) with 84 newspapers .

\begin{table}[htb]
\centering
\begin{tabular}{lrrrrrr}
\hline
              & Mean    &     SD & Median & Skew &   Min & Max \\
\hline
\#NP2023     &    2.04 &   2.99 &   1.00 & 11.4 &   0 & 84 \\
\#NP2018    &    2.29 &   3.69 &   1.00 & 12.9 &   0 & 107 \\   
%Population    &  102,790 & 329,958 &  25,735 & 13.8 &    75 & 10,098,052 \\  
Populat.    &  103K & 330K &  26K & 13.8 &    75 & 10098K \\  
\% Black      &    8.92 &  14.47 &   2.2 & 2.32 &   0.0 & 87.4 \\ 
\% Hisp       &    9.24 &  13.75 &   4.1 & 3.10 &   0.0 & 99.1 \\ 
Income     &   51583 &  13704 &  49888 & 1.29 & 20188 & 136268 \\ 
Age           &   41.29 &   5.41 &  41.20 & 0.08 &  21.7 & 67.00 \\
\hline
ln(pop)       &   10.27 &   1.49 &  10.16 &  0.28  & 4.32 & 16.13 \\
ln(black)     &    1.50 &   1.20 &   1.16 &  0.66  & 0 & 4.48 \\
ln(Hisp)      &    1.82 &   0.93 &   1.63 &  0.75  & 0 & 4.61 \\
\hline
\end{tabular}
    \caption{Descriptive statistics ($n=3,141$, K means times 1000)}
    \label{tab:desstat}
\end{table}

\begin{figure}[htb]
    \centering
    \includegraphics[width=.5\textwidth]{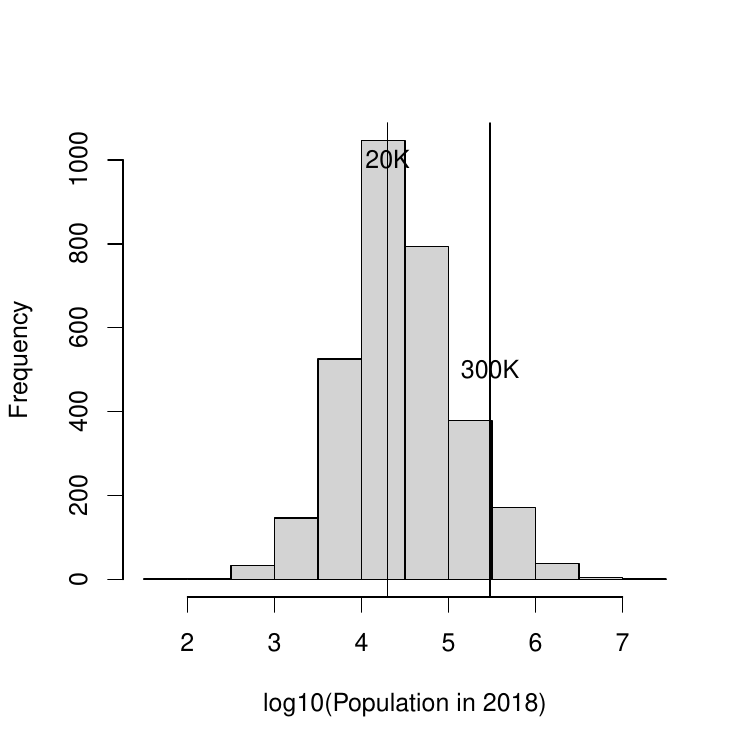}
    \caption{Histogram of the base-10 log population size in 2018}
    \label{fig:pophist}
\end{figure}

Our model allows for different slopes in the three population segments. Figure~\ref{fig:pophist} shows a histogram of population. There are 1,312 (41.8\%) counties with fewer than 20,000 people, 1,599 (50.9\%) counties with between 20,000 and 300,000 people, and 230 (7.3\%) counties with population greater than 300,000. These breaks were selected to be consistent with \citet{napoliAssessingLocalJournalism2018}.

\subsection{Model Validation I: Explanatory}
\label{ss:interpret}

We first present the Explanatory Model. Using generalized linear model (GLM) notation, the model is as follows:
{\small
\begin{verbatim}
NP2023 ~ popseg*(lnPop+age+lnHisp+lnBlack+HHincome),
\end{verbatim}
}
\noindent which allows for different slopes for all predictor variables in different population segments but does not include the lagged
number of newspapers. See
Table~\ref{tab:explainPartial} for likelihood ratio tests (LRT) of the interactions with population segment
and Table~\ref{tab:slopes}, columns 2--5, for the parameter estimates.
The LRT rows test the null hypothesis that the slopes for
particular variable are the same across segments, versus an alternative hypothesis that at least one segment has a different
slope. The magnitude of the LRT is the increase in model deviance if the interaction terms were dropped (versus having the
demographic variable as a main effect), and thus gives a measure of how important each interaction is, i.e., how much the slope
varies across population segments. We can conclude that the slopes for population vary greatly (LRT=159) across segments, the
slopes for age vary substantially (LRT=24) across segments, and the slopes for the other variables vary less across segments. We
will not do any post-hoc model selection and leave all interactions in the model, even though income is only borderline
significant ($P$=0.08968).

\begin{table}[htb]
\begin{tabular}{lrrrr}
\hline
               &Df &Deviance &      LRT & $P$-value \cr
\hline
<none>         &   &  2423.0 &          &           \cr
popseg:lnPop   & 2 &  2581.9 &  158.920 & $<0.0001$ \cr
popseg:age     & 2 &  2447.3 &   24.358 & $<0.0001$  \cr
popseg:lnHisp  & 2 &  2429.1 &    6.097 &   0.04744  \cr
popseg:lnBlack & 2 &  2429.8 &    6.783 &   0.03366  \cr
popseg:HHincome& 2 &  2427.8 &    4.823 &   0.08968  \cr
\hline
\end{tabular}
    \caption{Partial deviances for Explanatory Model}
    %Cpub2023 $\tilde$ popseg*(lnPop+age+lnHisp+lnBlack+HHincome)
    \label{tab:explainPartial}
\end{table}

\newcommand{\ten}{^{\dag}}
\begin{table*}[htb]
{\small
\begin{tabular}{l|rrrr|rrrr}
\hline
                       & \multicolumn{4}{c|}{\bf Explanatory Model} & \multicolumn{4}{c}{\bf Forecasting Model} \cr
                       & Estimate     & Std. Error  & $z$ value & $P$-value  &   Estimate     & Std. Error  & $z$ value & $P$-value \cr
\hline
(Intercept)            & $-3.708$    &    $0.4809$ & $ -7.711$ &  $<0.0001$ &    $-1.719$ &  $0.5101$ & $-3.370$ & $ 0.000752$ \cr
popseg<20K:ln(lagpub+1)&             &             &           &            &    $ 1.608$ &   $0.07229$ & $22.249$ & $<0.0001$ \cr
popseg<20K:lnPop       & $ 0.3577$   &   $0.03806$ & $  9.398$ & $< 0.0001$ &  $ 0.04240$ &   $0.04205$ &  $1.008$ & $ 0.313356$ \cr
popseg<20K:age         & $ 0.01092$  &  $0.005093$ & $  2.143$ & $0.032088$ & $ 0.002283$ &  $0.005543$ &  $0.412$ & $ 0.680397$ \cr
popseg<20K:lnHisp      & $ 0.008160$ &   $0.02805$ & $  0.291$ & $0.771117$ & $-0.002599$ &   $0.02889$ & $-0.090$ & $ 0.928311$ \cr
popseg<20K:lnBlack     & $-0.1250$   &   $0.02431$ & $ -5.140$ &  $<0.0001$ & $-0.002930$ &   $0.02595$ & $-0.113$ & $ 0.910096$ \cr
popseg<20K:HHincome    & $ 7.159\ten$& $2.257\ten$ & $  3.172$ & $0.001514$ &$0.5399\ten$ & $2.554\ten$ &  $0.211$ & $ 0.832566$ \cr
popseg20-300K          & $ 0.4712$   &    $0.5928$ & $  0.795$ & $0.426635$ &    $0.6230$ &    $0.6185$ &  $1.007$ & $ 0.313825$ \cr
popseg20-300K:ln(lagpub+1)&          &             &           &            &    $ 1.297$ &   $0.04456$ & $29.102$ & $<0.0001$ \cr
popseg20-300K:lnPop    & $ 0.3508$   &   $0.02765$ & $ 12.688$ & $<0.0001$  &  $ 0.01584$ &   $0.03010$ &  $0.526$ & $ 0.598722$ \cr
popseg20-300K:age      & $ 0.005068$ &  $0.003970$ & $  1.277$ & $0.201760$ & $ 0.001569$ &  $0.004062$ &  $0.386$ & $ 0.699417$ \cr
popseg20-300K:lnHisp   & $-0.03864$  &   $0.02377$ & $ -1.625$ & $0.104103$ & $-0.004181$ &   $0.02397$ & $-0.174$ & $ 0.861534$ \cr
popseg20-300K:lnBlack  & $-0.1232$   &   $0.01864$ & $ -6.610$ &  $<0.0001$ & $-0.005516$ &   $0.01967$ & $-0.280$ & $ 0.779150$ \cr
popseg20-300K:HHincome & $ 1.765\ten$& $1.515\ten$ & $  1.165$ & $0.243933$ &$0.2165\ten$ & $1.621\ten$ &  $0.134$ & $ 0.893740$ \cr
popseg300K+            & $-8.046$    &    $0.7427$ & $-10.834$ & $<0.0001$  &    $ 1.443$ &    $0.9084$ &  $1.588$ & $ 0.112285$ \cr
popseg300K+:ln(lagpub+1)&            &             &           &            &    $ 1.047$ &   $0.04701$ & $22.283$ & $<0.0001$ \cr
popseg300K+:lnPop      & $ 0.8865$   &   $0.03522$ & $ 25.169$ & $<0.0001$  &  $-0.05795$ &   $0.05369$ & $-1.079$ & $ 0.280498$ \cr
popseg300K+:age        & $ 0.04574$  &  $0.007150$ & $  6.397$ &  $<0.0001$ & $ 0.005171$ &  $0.007817$ &  $0.662$ & $ 0.508265$ \cr
popseg300K+:lnHisp     & $-0.1122$   &   $0.03990$ & $ -2.812$ & $0.004922$ &  $ 0.06493$ &   $0.04061$ &  $1.599$ & $ 0.109859$ \cr
popseg300K+:lnBlack    & $-0.02780$  &   $0.03392$ & $ -0.820$ & $0.412451$ &  $ 0.06100$ &   $0.03559$ &  $1.714$ & $ 0.086523$ \cr
popseg300K+:HHincome   & $ 5.208\ten$& $1.407\ten$ & $  3.702$ & $0.000214$ & $2.035\ten$ & $1.499\ten$ &  $1.358$ & $ 0.174554$ \cr
\hline
    Null deviance      & \multicolumn{4}{c|}{5947.6  on 3140 df} & \multicolumn{4}{c}{5947.6  on 3140df} \cr
Residual deviance      & \multicolumn{4}{c|}{2423.0  on 3123 df} & \multicolumn{4}{c}{550.5  on 3120df} \cr
\hline
\end{tabular}
}
    \caption{Estimated slopes for the \countmodel\ models ($\dag$ means multiply by $10^{-6}$)}
    \label{tab:slopes}
\end{table*}

The parameter estimates in columns 2--5 of Table~\ref{tab:slopes} give the estimated equations for the three segments and the $z$ values evaluate $H_0: \beta_j=0$ $(j=1,\ldots p)$ against a two-sided alternative. The predictor variables have different units and so the slopes are not directly comparable, whereas the $z$ values are unitless and can be compared. We discuss the different hypotheses:
\begin{itemize}
    \item H1: population has a very highly-significant ($P<.001$), strong, positive effect on the (log) expected count of
    newspapers for all segments, confirming H1. For all three segments the population $z$ values are the largest, and the effect of population is stronger for larger counties.
    \item H2: income also has highly-significant, positive effects in small and large population segments, but is not significantly different from 0 in medium-sized markets. We note that \citet{napoliAssessingLocalJournalism2018} also did not find significant income effects in their study of mid-size communities. 
    \item H3: the percentage of Blacks is very highly significant and negative in small and mid-size markets, but not significantly different from 0 in large markets. 
    \item H4: the percentage of Hispanics has a highly significant, negative effect in large markets, but is not significantly different from 0 in small and mid-size markets.
    \item H5: age has positive effects in large and small markets, but the effect is not significant in mid-size markets.
\end{itemize}

Thus, our model for mid-size markets largely confirms the results in \citet{napoliAssessingLocalJournalism2018}, while offering additional results for large and small markets. We consider the corroboration as a validation of our model.

%------------------------------------------------------------
\subsection{Forecasting model}

The Forecasting Model adds the logged, lagged number of newspapers as a predictor. Columns 6--9 of Table~\ref{tab:slopes} provide the estimated coefficients and
inference. See the next section for further validation of the model and discussion of the non-significant slopes.
%This discussion will be fairly technical and confined to its own subsection. 
This section proceeds with interpreting the forecasts from the model.

Readers may desire an $R^2$ value to assess how well the model predicts newspaper counts. Count variables are complicated because the variance of the DV increases with
its mean, and so simple $R^2$ values will be disproportionately affected by values with larger
variances/means. To address this issue, Pseudo-$R^2$ values have been
discussed \citep{heinzl2003pseudo} that are analogous to $R^2$ from linear models: 
\[ \mathrm{Pseudo}\;R^2 = 1 - \frac{\mathrm{residual\;deviance}}{\mathrm{null\;deviance}} = 
1- \frac{550.5}{5947.6} \approx 90.7\%. \]
In comparison, the Explanatory Model, which omitted the lagged newspaper count, has $R^2 = 1-2423/5947.6 = 59.3\%$, and so lagged newspapers explains a large amount of deviance, but the covariates alone also have substantial explanatory power.

A value of 90.7\% seems impressive, but it should be acknowledged that this is a global measure and that it is easy to explain the difference between, say, zero newspapers and dozens of newspapers. It will be more difficult to separate counties with zero versus one newspaper.
%------------------------------------------------------------
\subsection{Interpreting and Illustrating the 2028 Predictions}

\begin{figure*}[htb]
    \centering
    \includegraphics[width=.9\textwidth]{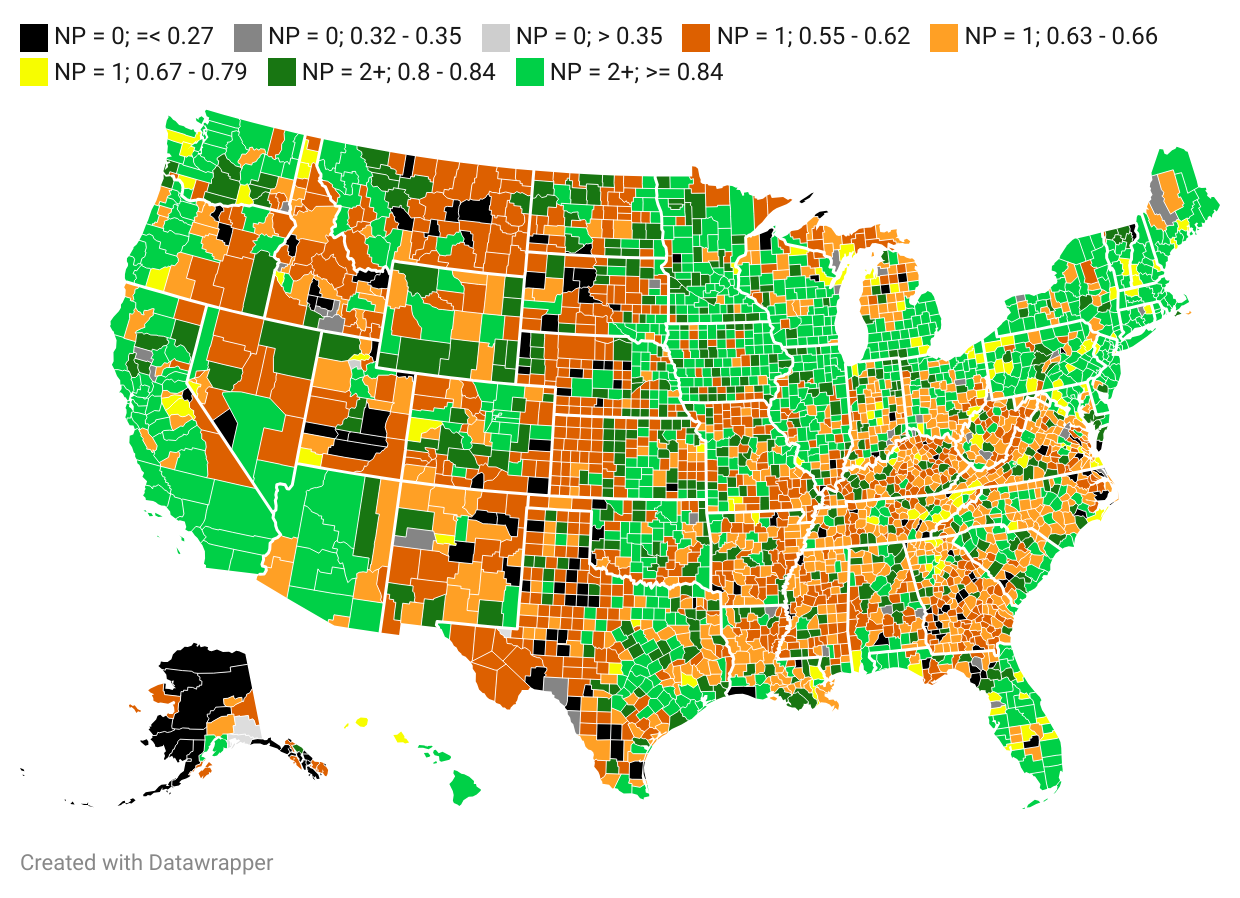}
    \caption{Comprehensive map showing predicted number of newspapers in 2028
    (numbers in legend give the probability of \underline{not} being a news desert)}
    \label{fig:MapV2}
\end{figure*}

We applied the Forecasting Model from Table~\ref{tab:slopes} (columns 6--9) to the current data, i.e., the number of newspapers in 2023 is now the lagged newspaper count and used market conditions from 2021\footnote{The most recent ACS five-year estimates are from 2021.} rather than 2018. The predictions from this model forecast the number of newspapers the county will have in 2028 and are shown on the ``comprehensive map'' in Figure~\ref{fig:MapV2}. The simplified version of the map shown on the News Desert website (Figure~\ref{fig:MapV1}) will be discussed in the next section.

\begin{figure*}[htb]
    \centering
    \includegraphics[width=.8\textwidth]{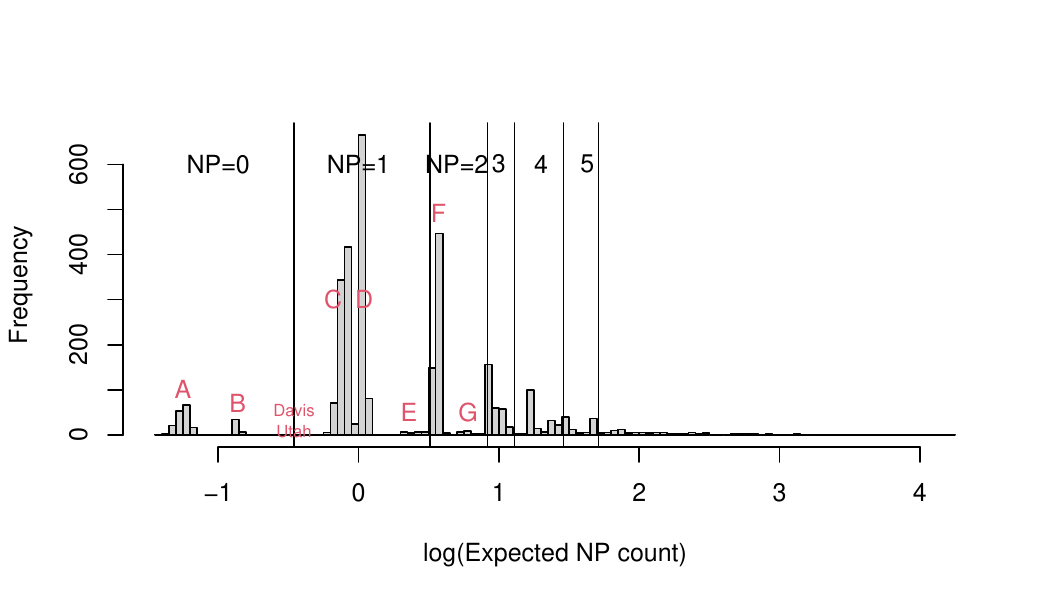}
    \caption{Histogram of log expected newspapers (linear predictor)}
    \label{fig:muhat}
\end{figure*}

%All predicted values are available in an Excel spreadsheet.\jwadd{(I don't think we need this sentence here?)} 
Figure~\ref{fig:muhat} shows a histogram of the log mean (expected) newspaper count from the Forecasting Model. Since Poisson regression uses a log link function, these are the values of \emph{linear predictors} $\hat{\eta}_i = \log(\mu_i) = \betab\myT\xb_i$, where $\hat{\mu}_i = e^{\hat{\eta}_i}$ are the forecasts. The histogram consists of many fairly-distinct modes, which correspond to the lagged number of newspapers in a market, as indicated by the labels across the top, e.g., ``NP=0''. Table~\ref{tab:muhatbycpub} shows the mean, minimum and maximum values of the predictions computed by the lagged newspaper count (for the future forecast, those in 2023). There is very little overlap. For example, among the complete news deserts where \#NP=0, the log expected count is between $-1.42$ and $-0.46$ with a mean of $-1.17$, which corresponds to the left-most mode in Figure~\ref{fig:muhat}. There is no overlap between this mode and the mode for counties with one newspaper in 2023. The vertical lines in Figure~\ref{fig:muhat} correspond to the maximum values of the linear predictors in Table~\ref{tab:muhatbycpub}, e.g., $-0.46$, 0.51, 0.92, \ldots. The histogram shows that there is variation within the modes, which is due to differences in market conditions, including the population segment. This variation is important because it means that not all news deserts are equally likely to remain so. The expected number of newspapers in five years ranges from 0.24 to 0.63, with a mean of 0.31. 

\begin{table*}[htb]
    \centering
    \begin{tabular}{c|rrr|rrr|rrr}
\hline
\# NP & \multicolumn{3}{c|}{log expected count} & \multicolumn{3}{c|}{Expected count} & \multicolumn{3}{c}{$\Prob($\#NP$>0$)}\\
\cline{2-10}
2023 & mean & min & max & mean & min & max  & mean & min & max\\
\hline
0 & $-1.17$ & $-1.42$ & $-0.46$ & 0.31 & 0.24 & 0.63 & 0.27 & 0.21 & 0.47 \\
1 & $-0.04$ & $-0.22$ & 0.51 & 0.97 & 0.80 & 1.66    & 0.62 & 0.55 & 0.81 \\
2 & 0.57 & 0.47 & 0.92 & 1.76 & 1.60 & 2.52          & 0.83 & 0.80 & 0.92 \\
3 & 0.97 & 0.90 & 1.11 & 2.63 & 2.47 & 3.04          & 0.93 & 0.92 & 0.95 \\
4 & 1.28 & 1.18 & 1.46 & 3.59 & 3.24 & 4.29          & 0.97 & 0.96 & 0.99 \\
5 & 1.50 & 1.35 & 1.71 & 4.49 & 3.86 & 5.54          & 0.99 & 0.98 & 1.00 \\
6 & 1.67 & 1.54 & 1.78 & 5.30 & 4.65 & 5.94          & 0.99 & 0.99 & 1.00 \\
7 & 1.83 & 1.75 & 1.91 & 6.25 & 5.73 & 6.74          & 1.00 & 1.00 & 1.00 \\
%8 & 1.92 & 1.79 & 2.05 & 6.84 & 5.97 & 7.77          & 1.00 & 1.00 & 1.00 \\
$\vdots$ & $\vdots$ & $\vdots$ & $\vdots$ & $\vdots$ & $\vdots$ & $\vdots$  & $\vdots$ & $\vdots$ & $\vdots$ \\
\hline
    \end{tabular}
    \caption{Summary statistics of predictions by the actual number of newspapers in 2023}
    \label{tab:muhatbycpub}
\end{table*}

\begin{figure*}[htb]
    \centering
    \includegraphics[width=.8\textwidth]{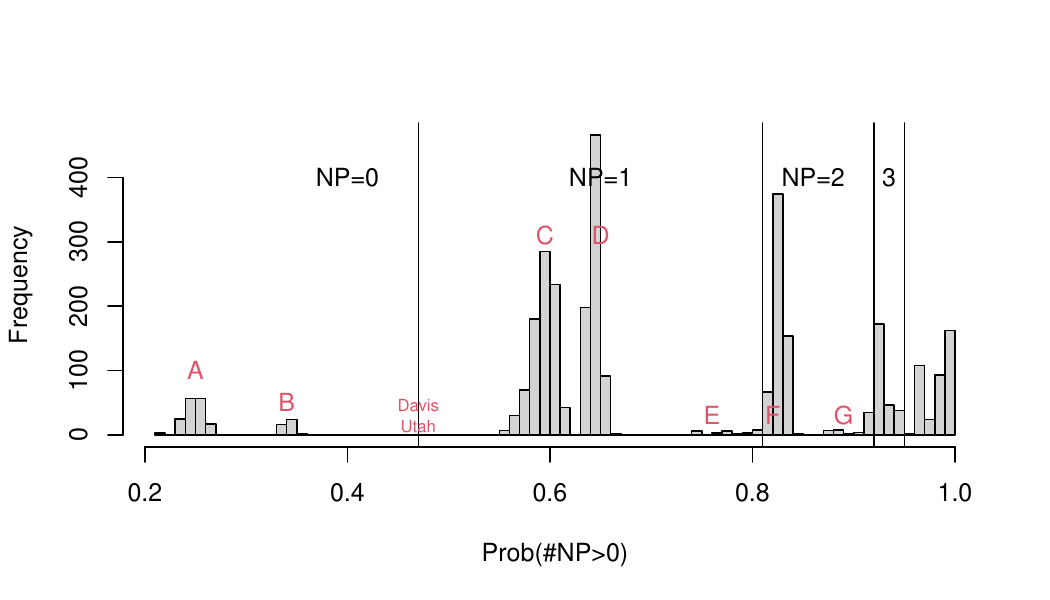}
    \caption{Histogram of the predicted probability of being a news desert (NP=0)}
    \label{fig:Pnd}
\end{figure*}

There is another way to view the predictions beyond expected counts and linear predictors (log expected counts). By invoking the Poisson assumption we can compute estimated probabilities of having some number of newspapers in the future. Recall that the probability mass function of a Poisson random variable $Y$ with parameter (mean) $\mu>0$ is 
\[ \Prob(Y=y) = \frac{e^{-\mu}\mu^y}{y!}, \quad y=0, 1, 2, \ldots.\] 
The probability of becoming a news desert is of particular interest:
\[ \Prob(Y=0) = \frac{e^{-\mu}\mu^0}{0!} = \frac{e^{-\mu}\cdot 1}{1} = e^{-\mu}. \]
Perhaps it is more intuitive to think of the probability of \emph{not} being a news desert, i.e., $\Prob(Y>0) = 1-e^{-\mu}$. Thus there is a one-to-one mapping from expected counts to the probabilities of being a news desert. For example, the probability that a county with estimated count 0.24 remains a news desert is $\Prob(Y=0) = e^{-0.24} \approx 79\%$, and therefore a 21\% chance of not being a news desert. The mean, minimum and maximum of the probabilities are provided in the last three columns of Table~\ref{tab:muhatbycpub} by lagged number of newspapers. For example, the news desert counties in 2023 have between a 21\% and 47\% of having at least one newspaper in the next five years. 
We use the variation around distinctive modes as the basis for our coloring scheme of map visualization. 
% This variation will be critical in constructing maps, because we will have to select cut points for showing different colors. 

Figure~\ref{fig:Pnd} shows a histogram of the predicted probabilities of \emph{not} being a news desert $\Prob(Y>0)$, which is simply a rescaled version of Figure~\ref{fig:muhat}. We emphasize that these forecasts are probabilities. A county with a 63\% chance of remaining a news desert has a 37\% chance of not being a news desert; some counties are more likely than others to remain news deserts. We now discuss some of the modes revealed on Figure~\ref{fig:Pnd}, which are depicted on the map in Figure~\ref{fig:MapV2}. 

%The histograms reveal some interesting ``outliers.'' 
\begin{itemize}
    \item Among the counties that were \textbf{news deserts} in 2023, there are two distinct modes and an outlier: 
    \begin{itemize}
        \item Mode A (black on map) has probabilities of not being a news desert between .27 and .21, and includes 160 newspapers that are very likely to remain a news desert.
        \item Mode B (gray on map) has probabilities between .33 and .35, and includes 42 newspapers that are somewhat less likely to remain a news desert. 
        \item The outlier (light gray on map) is Davis County, Utah (Salt Lake City suburb), and has a probability of .47 of not being a desert.
    \end{itemize}
    \item Among the counties with \textbf{one newspaper} in 2023, there are two somewhat distinct modes and a third group:
    \begin{itemize}
        \item Mode C (dark orange on map) has probabilities of not being a news desert between .55 and .62. There are 849 newspapers in this mode. These probabilities are substantial and the counties should be treated with high concern. These are called ``more at risk'' in Figure~\ref{fig:MapV1}.
        \item Mode D (orange on map) has probabilities of not being a news desert between .63 and .66. There are 758 newspapers in this mode. These probabilities are somewhat larger but still substantially less than 1. These are called ``less at risk'' in Figure~\ref{fig:MapV1}.
        \item Mode E (yellow on map) has probabilities of becoming a news desert between .73 and .81 and is not well-separated from mode F, with counties having two newspapers. There are 21 newspapers in this group. These probabilities are larger (relatively less likely to become a news desert) but still less than 1. These are called ``less at risk'' in Figure~\ref{fig:MapV1}.
    \end{itemize}
    \item Among counties with two newspapers in 2023, there is a natural break around a probability of 0.86.
    \begin{itemize}
        \item Mode F (dark green on map) has probabilities of not being deserts between .80 and .84. These are 606 two-newspaper counties that are somewhat more at risk.
        \item Mode G has probabilities between .86 and .92. There are 23 two-newspaper counties are safer.
    \end{itemize}
    \item There are 680 counties with more than 3 newspapers. They have probabilities between .92 and 1 of not being deserts.
\end{itemize}

%{\color{red} Measures for classification models like precision, recall and AUC may be more informative.}

%------------------------------------------------------------
\subsection{Additional Maps}
\label{ss:map}

We provide two additional versions of the map from the same set of predicted probabilities. Version 1 (``Barometer'') is the simplest version intended for the general public. Version 2 highlights counties that are notably over-/under-performing based on the difference between the actual and expected number of newspapers (residuals).

% We provide two versions of the map \jwadd{from the same set of predicted probabilities}. Version 1 has more color gradation and is intended for the academic audience. Version 2 is a simplified version intended for the general public. 
% Both maps are created from the same set of predicted probabilities.

\subsubsection{``Barometer'' Map from News Desert Website}

\begin{figure*}[htb]
    \centering
    \includegraphics[width=.9\textwidth]{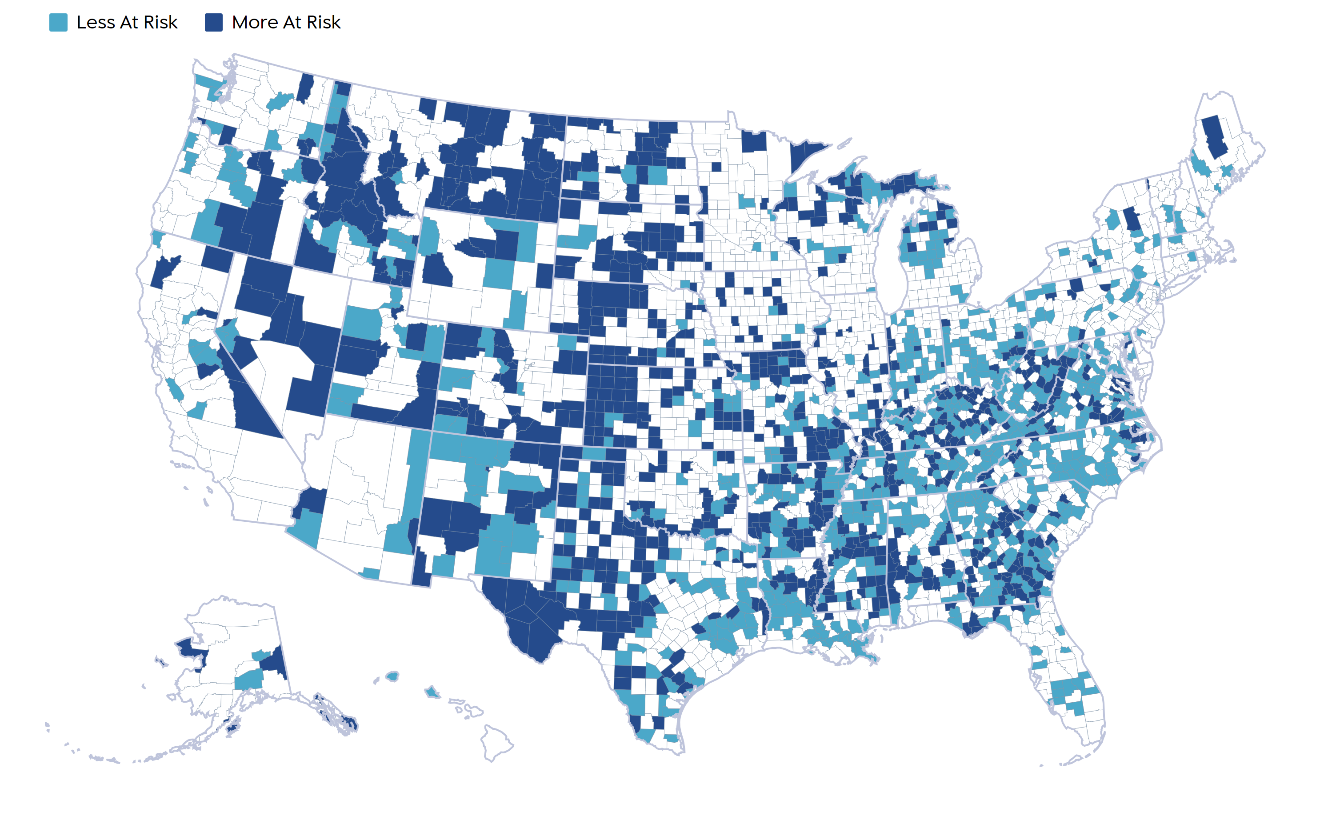}
    \caption{Counties currently with one newspaper and the forecast of whether they are more (mode C) or less (modes D and E) likely to become news deserts with no newspapers in five years.}
    \label{fig:MapV1}
\end{figure*}

Figure~\ref{fig:MapV1} was displayed on the \href{https://localnewsinitiative.northwestern.edu/}{News Desert website}. This section provides the details. We use the predicted values for 2028. This simplified version only shows counties with one newspaper and uses two colors. Dark blue shows counties that are more at risk, defined by Mode C in Figure~\ref{fig:Pnd}. Light blue shows counties that are less at risk, defined by modes D and E.

\subsubsection{Residual Map}

\begin{figure*}[htb]
    \centering
    \includegraphics[width=.9\textwidth]{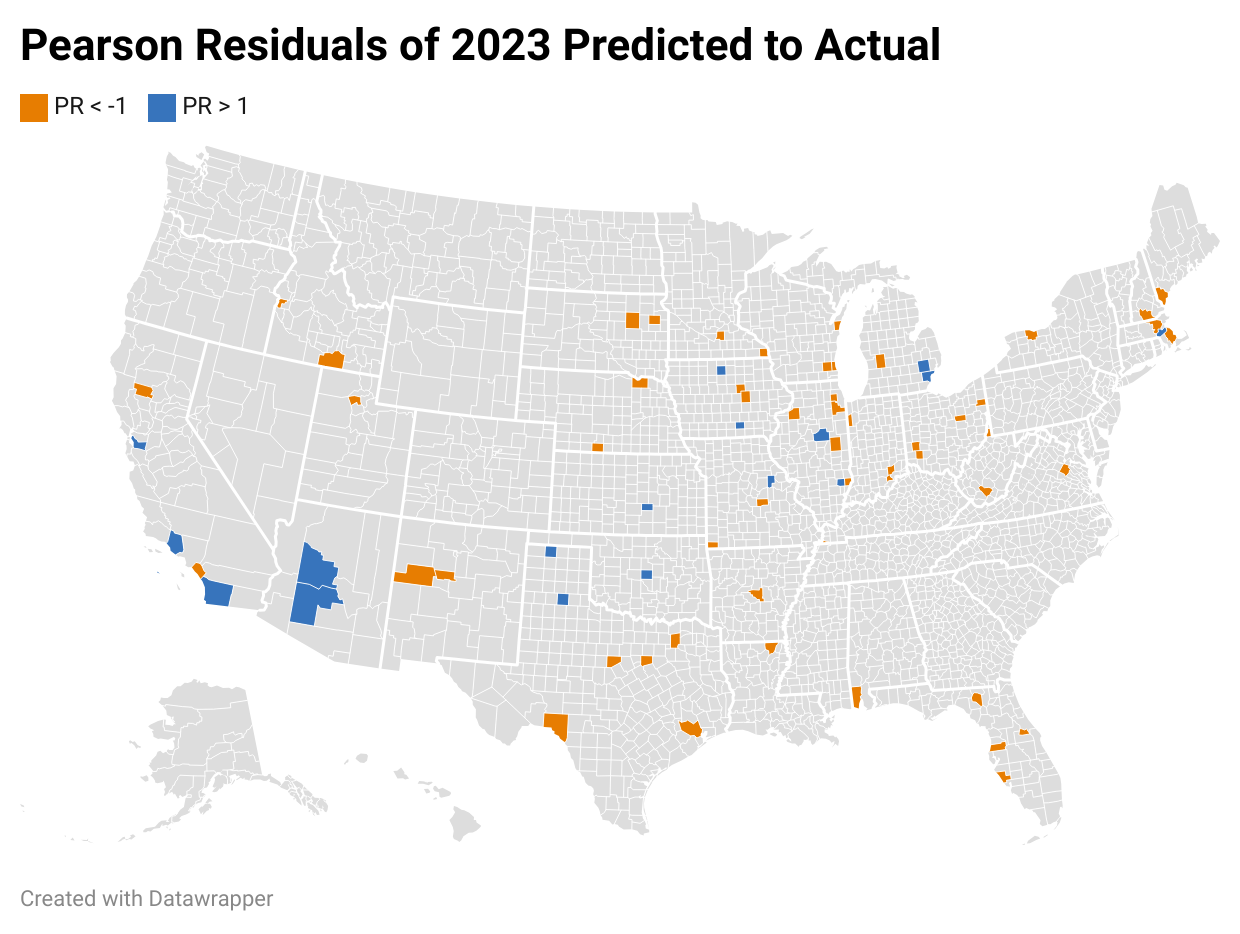}
    \caption{Map showing residuals}
    \label{fig:resids}
\end{figure*}

Figure~\ref{fig:resids} shows a map of the Pearson residuals from the Forecasting Model.
Residuals are defined as follows:
\[ r_i = \frac{y_i - \hat{\mu}_i}{\sqrt{\hat{\mu}_i}}. \]
%\[ r_i = \frac{y_i - \hat{\mu}_i}{\sqrt{\hat{\varphi}\hat{\mu}_i}}, \]
%where $\hat{\varphi} = 1/(n-p)\sum_{i=1}^n (y_i - \hat{\mu}_i)^2/\hat{\mu}_i$, is a scaling constant that adjust for over dispersion. 
They quantify what is \emph{not} explained by the model and can be thought of as a $z$ statistic: (observed value $-$ its mean) / standard deviation, since $\E(Y_i) = \hat{\mu}_i$ and $\sqrt{\V(Y_i)} =\sqrt{\hat{\mu}_i}$ is the standard deviation of $Y_i$.  

It is important to understand exactly what the residuals mean and their implications. Values near 0 indicate that the number
of newspapers is roughly what we would expect given the market conditions. The map in Figure~\ref{fig:resids} shows all
residuals between $-1$ and 1 as light grey. Positive values (residuals greater than 1 shown in blue) indicate counties that
are \textbf{over-performing} in that they have more newspapers than would be expected for the market conditions. Negative
values (residuals less than $-1$ shown in orange) indicate counties that are \textbf{under-performing}, with fewer newspapers than would be expected for their market conditions. 

Counties with large residuals, either negative or positive, should be studied further for several reasons. First, they may suggest additional covariates that should be included in the Forecasting Model. For example, suppose after further examination, counties with positive residuals have a certain type of ownership model while counties with negative residuals do not have that ownership model. This would suggest that ownership is an important factor that should be included in future iterations of the model and would reduce the magnitude of the residuals, thereby improving predictions and Pseudo-$R^2$. Doing in-depth interviews with journalists and publishers in counties with positive residuals---places that have healthier news ecosystems than one would expect given the market conditions---may produce insights for how to manage newspapers in difficult environments. 

%------------------------------------------------------------
\subsection{Model Validation II}
\label{ss:validationII}

This section is more technical and can be skipped by readers who are not interested. The Forecasting Model,
labeled the \textbf{Poisson linear} model in this section, did not use any post-hoc model selection.
This section attempts to answer two questions. As noted earlier, the lagged number of newspapers is the only significant variable in the model, with strong positive coefficients. The only variable that is even borderline significant is the percentage of Blacks ($P=0.0865$) in counties with large populations. The first question is, 
\hypoth{Q1}{Should we drop non-significant variables from the model?}
A slope being ``non-significant'' means that a 95\% confidence interval, which gives a window of values that the true slope could equal, includes the value 0, meaning the variable has no effect. Thus 0 is a plausible slope value, but it does not mean that the true slope is 0. The primary concern when building a forecasting model is whether including certain variables leads to overfitting. If some coefficient equals 0 then including it will cause the model to overfit and explain noise in the data. Q1 thus considers whether the Poisson linear model, which includes non-significant covariates, is too complex. 

The second question will consider the opposite question of whether the linear model is insufficiently complex.
We avoid problems with extreme skewness and influential observations by
applying basic log transformation, and consider the interactions with
population segment motivated by
\citet{napoliAssessingLocalJournalism2018}. However, our model has
not allowed for further nonlinearities or interactions. If such
nonlinearities exist, the Poisson linear model would be misspecified and the predictions
(forecasts) would be biased. 
\hypoth{Q2}{Are there additional nonlinear relationships?}

We answer both questions by first randomly selecting approximately 20\% of the counties (633) as a \textbf{test set}. The remaining 80\%
of counties (2508) will be used as the \textbf{training set}. We estimate five models using only the training set and then apply them to
the test set. This is a common procedure for evaluating machine-learning models \citep[\S5.1]{james2013introduction}. In particular, the
five models are:
\begin{itemize}
    \item \textbf{Poisson linear}: reestimate the Forecasting Model using only the training data.
    \item \textbf{AR(1)}:  Drop all market characteristics and include only the lagged number of newspapers as a predictor.
\end{itemize}

When a model overfits then dropping variables is one possible remedy, but there are other remedies under the names \emph{regularization} or \emph{shrinkage}. Ridge and lasso regression are examples. Rather than setting a coefficient to 0, such methods shrink the coefficient toward 0 to reduce the variable's influence on the predicted values and thereby the risk of overfitting. For large amounts of shrinkage the coefficients can equal 0 (or approach 0 in the case of ridge regression), but ridge and lasso offer alternatives beyond either having a variable in the model or not \citep{malthouse1999ridge}. The amount of shrinkage that is required to avoid overfitting is typically determined by cross validation (CV) \cite[\S6.2,6.6]{james2013introduction}. Ridge and lasso use different penalty terms to shrink the coefficients. 
\begin{itemize}
    \item \textbf{Lasso}: We re-estimated the Poisson linear model using a lasso shrinkage penalty term
    \citep[\S6.2]{james2013introduction} and 10-fold CV to select the shrinkage hyperparameter. %We use R's \texttt{glmnet} library.
    \item \textbf{Ridge}: We re-estimated the Poisson linear model using a ridge shrinkage penalty term
    \citep[\S6.2]{james2013introduction} and 10-fold CV to select the shrinkage hyperparameter. %We use R's \texttt{glmnet} library.
    \item \textbf{GBM}: gradient boosting machines are a powerful machine-learning ``black box'' method \cite[\S8.2]{james2013introduction} that automatically detect nonlinearities and interactions. We use R's \texttt{gbm} library with the default learning rate, an interaction depth of 2 (so that interactions are allowed), and 10-fold CV to select the number of trees.
\end{itemize}
Having estimated each of the models using only the training data, including the selection of the shrinkage parameter for lasso and ridge and the number of trees for GBM,  we apply the estimated models to the common test set and compare the accuracy of the predictions. To answer Q1, if the Poisson linear model outperforms AR(1) on the test set then we conclude that the market characteristics are contributing to the predictions and should remain in the model. The lasso and ridge models further check whether the Poisson linear model can predict better by shrinking the coefficients toward 0. We answer Q2 with the GBM model, which, in theory since GBMs are universal approximators, should provide an upper bound for how well we can predict with the variables we have in the model. If GBM substantially outperforms the Poisson linear model then there must be nonlinearities that have not been modeled and we should add terms to the Poisson linear model to improve it. Alternatively, if the two models have comparable accuracy then it is unlikely that the Poisson linear can be improved through transformations (although additional predictor variables could improve both models).

Table~\ref{tab:testDev} gives the deviance values for the test set and computes Pseudo-$R^2$ values, e.g., the Poisson linear model has $R^2 = 1-114.509/1111.304$. To answer Q1, the Poisson linear model outperforms the AR(1) and lasso models. Thus, even though the demographics are not statistically significant, the cross-validated lasso model suggests that their coefficients are non-zero, albeit small in magnitude. Moreover, shrinking them with ridge or lasso does not improve the predictions. The predicted values are largely determined by the lagged number of newspapers in the county, but, importantly, the demographic variables offer refinements that are not simply explaining noise in the data. In response to Q2, the Poisson linear model performs nearly as well as GBM, suggesting that additional transformations will not improve the model.
% Figure~\ref{fig:lasso} plots the cross-validated deviance against the shrinkage parameter. The cross-validated deviance is minimized when $\log(\lambda)=-6.805$, i.e., $\lambda=0.0011 \approx 0$. A value of $\lambda=0$ corresponds to no shrinkage, which would be the values in Table~\ref{tab:forecast}. The function shown in Figure~\ref{fig:lasso} is very flat, suggesting that the deviance would change very little as $\lambda\rightarrow 0$ (i.e., no shrinkage). For simplicity, we decided to use the unshrunk values from Table~\ref{tab:forecast}. Furthermore, the correlation between the predicted values from the lasso and unshrunk models is 0.99948, suggesting there is little practical difference between the two models. 

\begin{table}[htb]
    \centering
    \begin{tabular}{l|r|r}
    \hline
    Model & Deviance & Pseudo-$R^2$ \\
    \hline
    Null (intercept)& 1111.304 & 0\% \\
    Poisson linear  & 114.509 & 89.70\% \\
    AR(1) & 127.870 & 88.49\% \\
    Lasso     & 115.542 & 89.60\% \\
    Ridge & 133.208 & 88.01\% \\
    GBM       & 112.182 & 89.91\% \\
    \hline
    \end{tabular}
    \caption{Test set deviances and Pseudo-$R^2$ values}
    \label{tab:testDev}
\end{table}
%\begin{figure}
%    \centering
%    \includegraphics[width=.8\textwidth]{lasso.pdf}
%    \caption{Plot of 10-fold cross validated deviance versus the amount of lasso shrinkage}
%    \label{fig:lasso}
%\end{figure}

%------------------------------------------------------------
%\subsection{The counties most at risk}
%\label{ss:atrisk}
%
%We have been asked which counties are most at risk? We interpret the question to mean counties that currently have at least
%one newspaper, which are most likely to become complete news deserts with zero newspapers in the next five years? Table~\ref{tab:muhatbycpub} shows that the newspapers that are most at risk are those with one newspaper in 2023, since there is almost no overlap in predicted values with counties having two newspapers. There is substantial variation in the expected counts, ranging from 0.80 to 1.66, and we must select a cutoff to separate those most at risk versus those with lower risk. Figure~\ref{fig:muhat} reveals two modes in this range, which could be considered a natural division. Scanning through the list of counties sorted by predicted value, those most at risk tend to be in small counties (population less than 20K), although there are few few larger counties with lower expected counts.

%===================================================================
\section{Discussion}

We created and validated predictive models that anticipate where news deserts are likely to arise in five years, and identified some counties that are currently news deserts that are less likely to remain so, representing opportunities. 
The most important predictor of news deserts is the lagged number of newspapers, which mediates the relationship between market conditions and the number of newspapers. After dropping the lagged number of newspapers from the model to avoid ``blocking the pipe'' between market conditions and the newspaper counts, our hypotheses are largely supported. The variables are roughly discussed in order of importance, as determined by the absolute magnitude of the $z$ statistics. \textbf{Population size} is the strongest predictor, especially in large counties, with a positive effect on newspaper counts. The percentage of \textbf{Blacks} in the county has large negative effects on newspapers in small- and mid-sized counties, but the effect is not significant in large counties. \textbf{Age} has a strong positive effect in large counties, a weaker but significant effect in small counties, and no significant effect in mid-sized counties. \textbf{Income} has a positive effect on newspapers in large and small counties, but not in mid-sized counties. The percentage of \textbf{Hispanics} has a significant, negative effect on newspapers in large counties, but no significant effect in small- or mid-sized counties.

We offer several contributions to the literature. The work that is closest to
ours is \citet{napoliAssessingLocalJournalism2018}, although their purpose
was explanatory rather than making forecasts. Our approach extends theirs to
all counties in the US rather than only the mid-size communities.  We are the first to include lagged newspaper counts as a predictor and use age as a predictor. We found significant effects for income and the percentage of Blacks, in addition to corroborating their findings on population.

%------------------------------------------------------------
%\ecmadd{Policy implications}

%------------------------------------------------------------
\subsection{Future research}
\label{ss:futureResearch}

Some counties, especially those with large populations, are heterogeneous and comprised of many smaller communities. Some
communities have healthy news ecosystems with many newspapers, and adjacent communities within the same county may have
none. For example Cook County, Illinois has the most newspapers of any county in the US with a wide variety of different
types of newspapers, yet there are many neighborhoods without any coverage. We are only able to map newspapers
to counties and not to lower geographic levels such as five-digit postal zip codes. Obtaining lower-level data should
be a priority in the future.

Our dependent variable could be refined in different says. First, it would be desirable to include other types of media such as broadcast, radio, and ethnic media. Second, having a newspaper in a geographic unit does not guarantee that the unit receives coverage. Abernathy and others have warned of ghost newspapers, where the newspaper exists but has few reporters and is filled with stories from various wire services. It would be desirable to use natural language processing to automate \citet{napoliAssessingLocalJournalism2018} and monitor how much coverage each geographic unit receives, which could be used as the dependent variable.

There are opportunities to include additional predictor variables in the model. The number of existing newspapers
is the most important predictor, and we theorize that this could be due to brand equity. Future research could investigate
this explanation by including measures of brand equity. One simple measure could be the age of the brand. The ownership model
of the existing papers, e.g., privately held, hedge fund, etc., may affect the health of the newspaper and have important
policy implications for anti-trust. Another policy question concerns broadband infrastructure, which affects the ability of a
newspapers to shift from print to digital. For example, does government funding of broadband in rural areas reduce the chance
of a news desert? Some newspapers are shifting to non-profit business models, and the level of philanthropic support in a
newspaper's home market could be a critical predictor of the success of this model. The level of civic participation has been
shown to relate to news consumption \citep{ksiazekNewsseekersAvoidersExploring2010a}, and could be a good predictor. The level
of collaboration between newspapers and the sharing of resources is also desirable to include.

While data sources for many of these factors exist, are readily available and would likely improve our predictions, we have not used them in our model because the causality and endogeneity become very complicated. For example, consider broadband penetration, which is likely a consequence of other predictor variables in the model such as population density, as well as possibly income, and race. Some variables are endogenous, such as civic participation---a county with high civic engagement may be more likely to support a local newspaper, but the existence of a newspaper may cause residents to become more civically engaged. Models where the primary goal is explanation will be especially difficult to specify and interpret. 

Our work has uncovered interactions with population segments and some demographics that are associated with newspapers in some market sizes are not significant in others. We are unaware of theorization around these differences, yet understanding the reasons could help reduce the urban-rural divide.

Having developed this methodology for forecasting future news deserts, it would be desirable to gather data from other countries. This would allow us to evaluate the external validity of the model.

Finally, research is needed for acting on the predictions from this model. It is desirable to create a playbook for different types of counties. For example, among the counties that have a newspaper and have a high probability of losing it (mode C), what are the best practices for success? Within this group there may be additional types, e.g., those located in suburbs or exurbs may need different strategies than those in small, remote cities. Those located in areas with pockets of wealth may require different strategies than those in very poor areas. This may require in-depth interviews with media managers to understand successful business models.

%===================================================================
\section*{Acknowledgement}
We thank Tim Franklin, Penelope Abernathy and George Stanley for helpful suggestions in formulating
the model and for access to the news desert data.

%===================================================================
%\bibliographystyle{tfcad}
\bibliographystyle{ACM-Reference-Format}
\bibliography{refs.bib}

%===================================================================
\appendix
\section{Appendix}
The appendix provides additional models evaluating the robustness of the findings. 

\subsection{Replace income with SES}
\label{ss:ses}

SES is a composite variable and represents the first principal component of household income. Income and educational attainment are indicators of socioeconomic status, which can be highly correlated. Therefore, we examined underlying factor structure among three log-transformed Census variables: percent of population with bachelor’s degree and higher, percent of population with less than high school education, and median household income. The scree plot and eigenvalue indicate existence of one higher-level factor. We conducted principal component analysis to derive a single factor variable named socioeconomic status (SES). 

Table~\ref{tab:predSES} gives the estimates of a Poisson model replacing income with SES. The Pseudo-$R^2$ values are nearly equal (0.90749 versus 0.90744). The correlation between the linear predictions ($\hat{\eta}$) between the SES and income model is 0.9999275, indicating essentially identical predictions. 

\begin{table}[htb]
    \centering
    {\small
    \begin{verbatim}
                            Estimate Std. Error z value Pr(>|z|)  
(Intercept)               -1.6685688  0.4823957  -3.459 0.000542
popseg20-300K              0.6554298  0.6108765   1.073 0.283301
popseg300K+                1.4673588  0.8874128   1.654 0.098224
popseg<20K:ln(lagpub+1)    1.6052674  0.0724554  22.155 <2e-16
popseg20-300K:ln(lagpub+1) 1.2949427  0.0446352  29.012 <2e-16
popseg300K+:ln(lagpub+1)   1.0523893  0.0466023  22.582 <2e-16
popseg<20K:lnPop           0.0429556  0.0420869   1.021 0.307425  
popseg20-300K:lnPop        0.0079665  0.0318614   0.250 0.802560  
popseg300K+:lnPop         -0.0611809  0.0536328  -1.141 0.253979  
popseg<20K:age             0.0017383  0.0055352   0.314 0.753487  
popseg20-300K:age          0.0016370  0.0040564   0.404 0.686536  
popseg300K+:age            0.0055003  0.0077692   0.708 0.478964  
popseg<20K:lnHisp         -0.0026641  0.0289514  -0.092 0.926683  
popseg20-300K:lnHisp      -0.0006905  0.0245398  -0.028 0.977552  
popseg300K+:lnHisp         0.0836949  0.0451254   1.855 0.063636
popseg<20K:lnBlack        -0.0037701  0.0262886  -0.143 0.885966  
popseg20-300K:lnBlack     -0.0023205  0.0198961  -0.117 0.907151  
popseg300K+:lnBlack        0.0568380  0.0350717   1.621 0.105099  
popseg<20K:SES18           0.0061308  0.0319523   0.192 0.847841  
popseg20-300K:SES18        0.0150594  0.0240647   0.626 0.531453  
popseg300K+:SES18          0.0420962  0.0361639   1.164 0.244407  
    Null deviance: 5939.38  on 3138  degrees of freedom
Residual deviance:  549.45  on 3118  degrees of freedom
\end{verbatim}
}
    \caption{Prediction model replacing household income with a composite measure of socioeconomic status (SES)}
    \label{tab:predSES}
\end{table}

\subsection{Replace income with poverty level}
\label{ss:poverty}

The percentage of households in poverty accounts for the cost of living in an area, e.g., incomes that are below the poverty level in large cities may be above the poverty level in a rural area. The percentage of household in poverty, however, does not account for the upper end of the income distribution. Table~\ref{tab:predPoverty} gives the estimates. The Pseudo-$R^2$ values are nearly equal (0.9072435 versus 0.90744), the poverty model slightly worse in the fourth decimal place. The correlation between the linear predictions ($\hat{\eta}$) between the SES and income model is 0.9999203, indicating essentially identical predictions. 

\begin{table}[htb]
    \centering
    {\small
    \begin{verbatim}
                            Estimate Std. Error z value Pr(>|z|)  
(Intercept)               -1.6585841  0.4904385  -3.382  0.00072
popseg20-300K              0.5164064  0.6247865   0.827  0.40850  
popseg300K+                1.6179079  0.8948851   1.808  0.07061
popseg<20K:ln(lagpub+1)    1.6052447  0.0743828  21.581  <2e-16
popseg20-300K:ln(lagpub+1) 1.2975758  0.0445912  29.099  <2e-16
popseg300K+:ln(lagpub+1)   1.0578253  0.0464545  22.771  <2e-16
popseg<20K:lnPop           0.0429368  0.0422767   1.016  0.30981  
popseg20-300K:lnPop        0.0191363  0.0291864   0.656  0.51204  
popseg300K+:lnPop         -0.0616380  0.0536291  -1.149  0.25042  
popseg<20K:age             0.0016911  0.0056449   0.300  0.76449  
popseg20-300K:age          0.0018034  0.0041609   0.433  0.66472  
popseg300K+:age            0.0049990  0.0081432   0.614  0.53929  
popseg<20K:lnHisp         -0.0033357  0.0289454  -0.115  0.90825  
popseg20-300K:lnHisp      -0.0040492  0.0239100  -0.169  0.86552  
popseg300K+:lnHisp         0.0648928  0.0413603   1.569  0.11666  
popseg<20K:lnBlack        -0.0046426  0.0261496  -0.178  0.85908  
popseg20-300K:lnBlack     -0.0082742  0.0204179  -0.405  0.68530  
popseg300K+:lnBlack        0.0513554  0.0363358   1.413  0.15755  
popseg<20K:poverty        -0.0004684  0.0044036  -0.106  0.91528  
popseg20-300K:poverty      0.0009966  0.0039076   0.255  0.79869  
popseg300K+:poverty       -0.0030956  0.0062896  -0.492  0.62259  
    Null deviance: 5939.38  on 3138  degrees of freedom
Residual deviance:  550.92  on 3118  degrees of freedom
\end{verbatim}
}
    \caption{Prediction model replacing household income with the percentage of households below the poverty level}
    \label{tab:predPoverty}
\end{table}

\subsection{Add RUCC}
\label{ss:rucc}

We add RUCC as a main effect, which will introduce multicollinearity with population and the population segment. Table~\ref{tab:predRUCC} gives the estimates. The Pseudo-$R^2$ values are nearly equal (0.9077038 versus 0.90744), with the addition of RUCC improving $R^2$ slightly in the fourth decimal place. The correlation between the linear predictions ($\hat{\eta}$) between the SES and income model is 0.9997776, indicating essentially identical predictions. 

\begin{table}[htb]
    \centering
    {\small
    \begin{verbatim}
                            Estimate Std. Error z value Pr(>|z|)  
(Intercept)               -1.807e+00  5.811e-01  -3.109  0.00187
popseg20-300K              5.632e-01  6.901e-01   0.816  0.41439
popseg300K+                1.547e+00  9.745e-01   1.587  0.11244  
factor(RUCC13)2           -3.077e-03  4.782e-02  -0.064  0.94870  
factor(RUCC13)3            2.136e-02  5.798e-02   0.368  0.71261  
factor(RUCC13)4            4.038e-02  6.811e-02   0.593  0.55333  
factor(RUCC13)5            6.643e-02  9.028e-02   0.736  0.46184  
factor(RUCC13)6            2.799e-02  6.339e-02   0.442  0.65882  
factor(RUCC13)7            4.492e-02  6.877e-02   0.653  0.51357  
factor(RUCC13)8            5.460e-02  8.886e-02   0.614  0.53890  
factor(RUCC13)9            4.564e-02  8.712e-02   0.524  0.60039  
popseg<20K:ln(lagpub+1)    1.601e+00  7.270e-02  22.018  <2e-16
popseg20-300K:ln(lagpub+1) 1.295e+00  4.487e-02  28.864  <2e-16
popseg300K+:ln(lagpub+1)   1.048e+00  4.732e-02  22.143  <2e-16
popseg<20K:lnPop           4.962e-02  4.973e-02   0.998  0.31836  
popseg20-300K:lnPop        2.404e-02  3.389e-02   0.709  0.47809  
popseg300K+:lnPop         -5.898e-02  5.603e-02  -1.053  0.29252  
popseg<20K:age             1.831e-03  5.581e-03   0.328  0.74288  
popseg20-300K:age          1.722e-03  4.119e-03   0.418  0.67597  
popseg300K+:age            5.172e-03  7.818e-03   0.662  0.50823  
popseg<20K:lnHisp         -2.685e-03  2.946e-02  -0.091  0.92738  
popseg20-300K:lnHisp      -6.221e-03  2.414e-02  -0.258  0.79663  
popseg300K+:lnHisp         6.520e-02  4.083e-02   1.597  0.11030  
popseg<20K:lnBlack        -2.632e-03  2.615e-02  -0.101  0.91981  
popseg20-300K:lnBlack     -1.751e-03  2.009e-02  -0.087  0.93056  
popseg300K+:lnBlack        6.063e-02  3.606e-02   1.681  0.09271
popseg<20K:HHincome        7.676e-07  2.594e-06   0.296  0.76734  
popseg20-300K:HHincome     7.241e-07  1.797e-06   0.403  0.68693  
popseg300K+:HHincome       2.003e-06  1.580e-06   1.268  0.20480  
    Null deviance: 5939.38  on 3138  degrees of freedom
Residual deviance:  548.18  on 3110  degrees of freedom
\end{verbatim}
}
\caption{Prediction model adding RUCC as a predictor}
    \label{tab:predRUCC}
\end{table}

\end{document}